\documentstyle[11pt,newpasp,twoside,epsf]{article}
\newcommand{\mlapm}{\texttt{MLAPM}}
\def\nbody{$N$-body}
\def\zform{$z_{\rm form}$}
\def\hkpc{$h^{-1}{\ }{\rm kpc}$}
\def\hMpc{$h^{-1}{\ }{\rm Mpc}$}
\def\hMsun{$h^{-1}{\ }{\rm M_{\odot}}$}
\def\LCDM{$\Lambda$CDM}

\begin{document}
\title{Satellite galaxies in cosmological dark matter halos}
\author{Stuart P.~D. Gill, Alexander Knebe, Brad K Gibson}
\affil{Centre for Astrophysics \& Supercomputing, Swinburne University,
Australia}


\begin{abstract}
We present preliminary results from a series of high-resolution
\nbody\ simulations that focus on 8 dark matter halos each of order a
million particles within the virial radius. We follow the time
evolution of hundreds of individually tracked satellite galaxies and
relate their physical properties to the differing halo environmental
conditions. Our main science driver is to understand how satellite
galaxies lose their mass and react to tidal stripping. Unlike previous
work our results are performed in a fully self-consistent cosmological
context. The preliminary results demonstrate that while environment
may vary considerably with respects to formation time and richness of
substructure, the satellites evolve similarly.
\end{abstract}


\section{The Simulations}

Four standard  \LCDM\ cosmological simulations with side length 64\hMpc\ were run at low mass resolution ($128^3$ particles), with 8 halos selected sampling a range of environments. In this contribution we present results for our oldest (8.3 Gyrs) and youngest (3.4 Gyrs) dark matter halos, halo \# 1 and \# 8 respectively.  These halos were re-simulated at 64 times higher mass resolution giving an effective mass per particle of $10^8$ \hMsun, and a force resolution of approximately 1 \hkpc.  The \nbody\ simulations were carried out using the publicly available adaptive mesh refinement code \mlapm\ (Knebe, Green~\& Binney 2001). The code solves Poisson's equation on a hierarchy of nested grids: the whole computational volume is covered by one cubic domain grid whereas refined regions are of arbitrary shape and adjusted to the actual density field at each major step in order to follow the real distribution of particles at all times.  An example of \mlapm\ in action is shown in the left panel of Figure 1. These two images show the density field of our youngest halo superimposed by the refinement grids. The top image shows the 5th refinement level with grid resolution $4096^3$, the bottom the 6th level of refinement.  As the grids follow the over-densities in the simulations they by definition encompass the satellites.  Hence, we can build a hierarchy of nested isolated \mlapm\ grids into a "grid tree", and then generate a list of potential satellite centers  by treating the densest grid at the end of each branch as a "potential" halo center. We further check the validity of the center by assigning physical properties to the adjacent radial density distribution. The details can be found elsewhere (Gill, Knebe~\& Gibson 2003).
 
\begin{figure}
\vspace{88mm}
\includegraphics{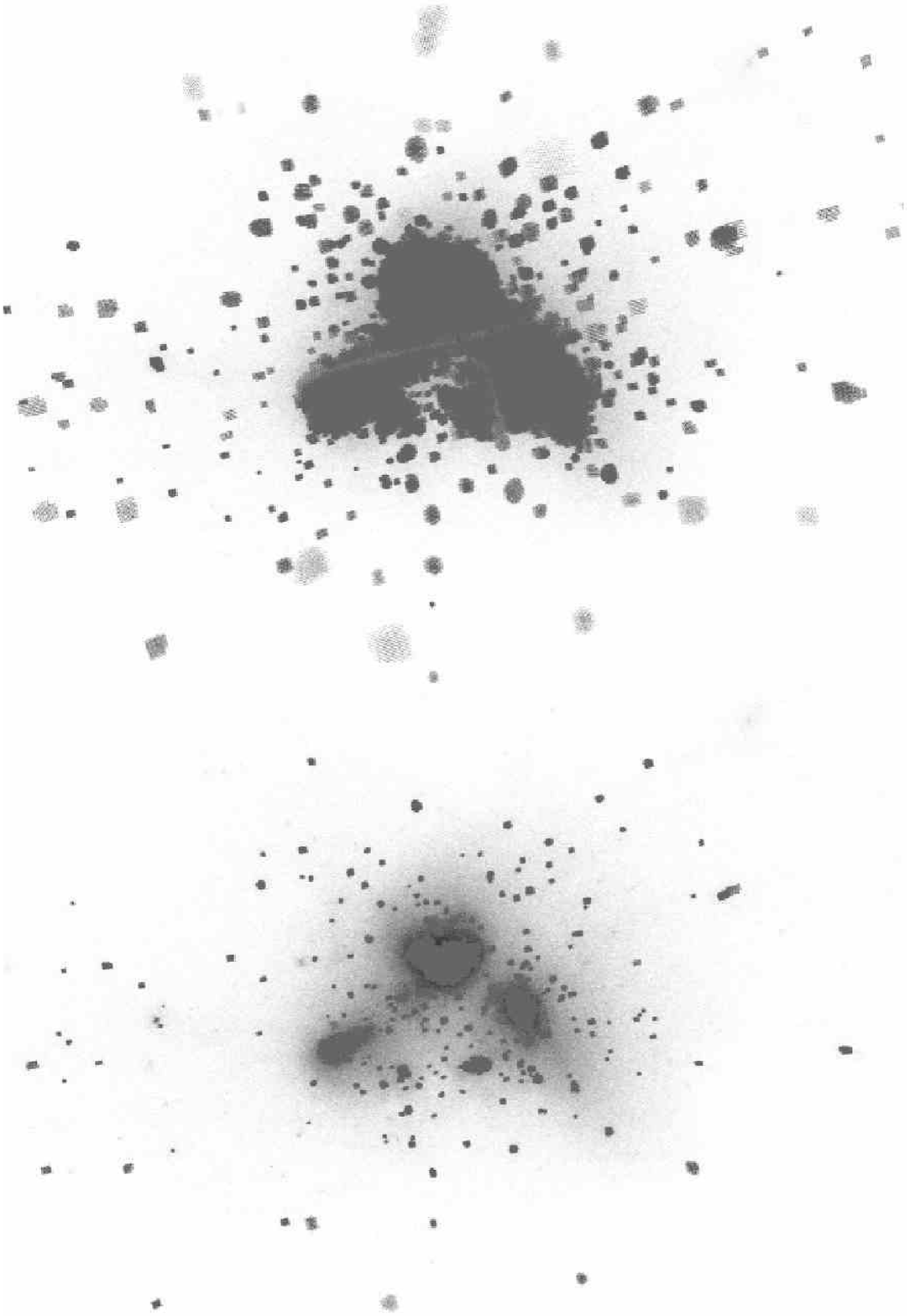}
\includegraphics{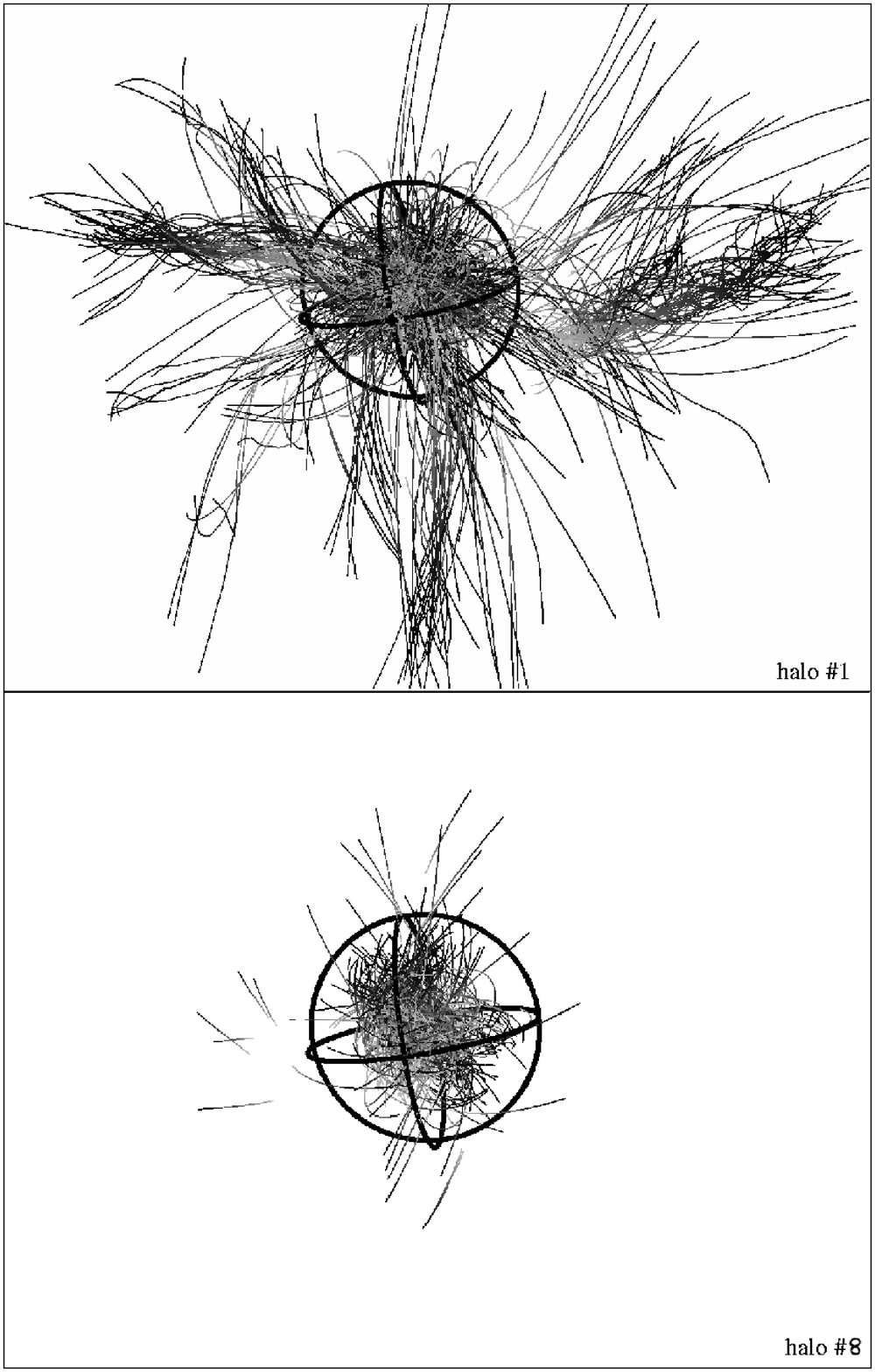}
   \caption{\footnotesize {(Left Panel) Consecutive refinement levels of \mlapm 's grid structure for halo \# 8 superimposed upon the density projection of the particle distribution. (Right Panel) Orbits of all objects in the vicinity of the host halo; orbital lines graduate from dark at \zform\ time to light $z=0$.}}
\end{figure}

\section{Dark Matter Halo Environment}

Following Lacey \& Cole (1994) the formation time of the halo \zform\ is defined to be the time when the halo contains half of its present day mass. From \zform\ to z=0 we track all satellites in the vicinity of the host halo. The number of satellites within the virial radius of our 8 halos (of mass $> 2 \times 10^{10}$ \hMsun ) ranges from $50 - 250$. The right panel of Figure 1 shows the orbits of the tracked satellites for two sample host halos highlighting the differing halo environments. The top image shows a halo developing in a rich region fed by three filaments, where as the figure directly below shows a host halo in a relatively isolated region that saw early rapid collapse.



\section{Satellite Disruption}

Figure 2 shows the number of satellites within a 2 \hMpc\ sphere (physical coordinates), as a function of redshift normalized by the number of satellites within the sphere at redshift \zform. The thin line (always on top) shows the total number of satellites within that sphere, while the bottom (thick) line shows the number of "disrupted" satellites subtracted from the number of satellites at \zform. The rise in the thin lines from \zform to z=0 reflects the richness of the immediate environment around the halo. Halo \#1 is in a particularly rich "filament fed region". These filaments are a source of significant satellite infall (recall the right panel of Figure 1).

\begin{figure}
\vspace{75mm}
\includegraphics{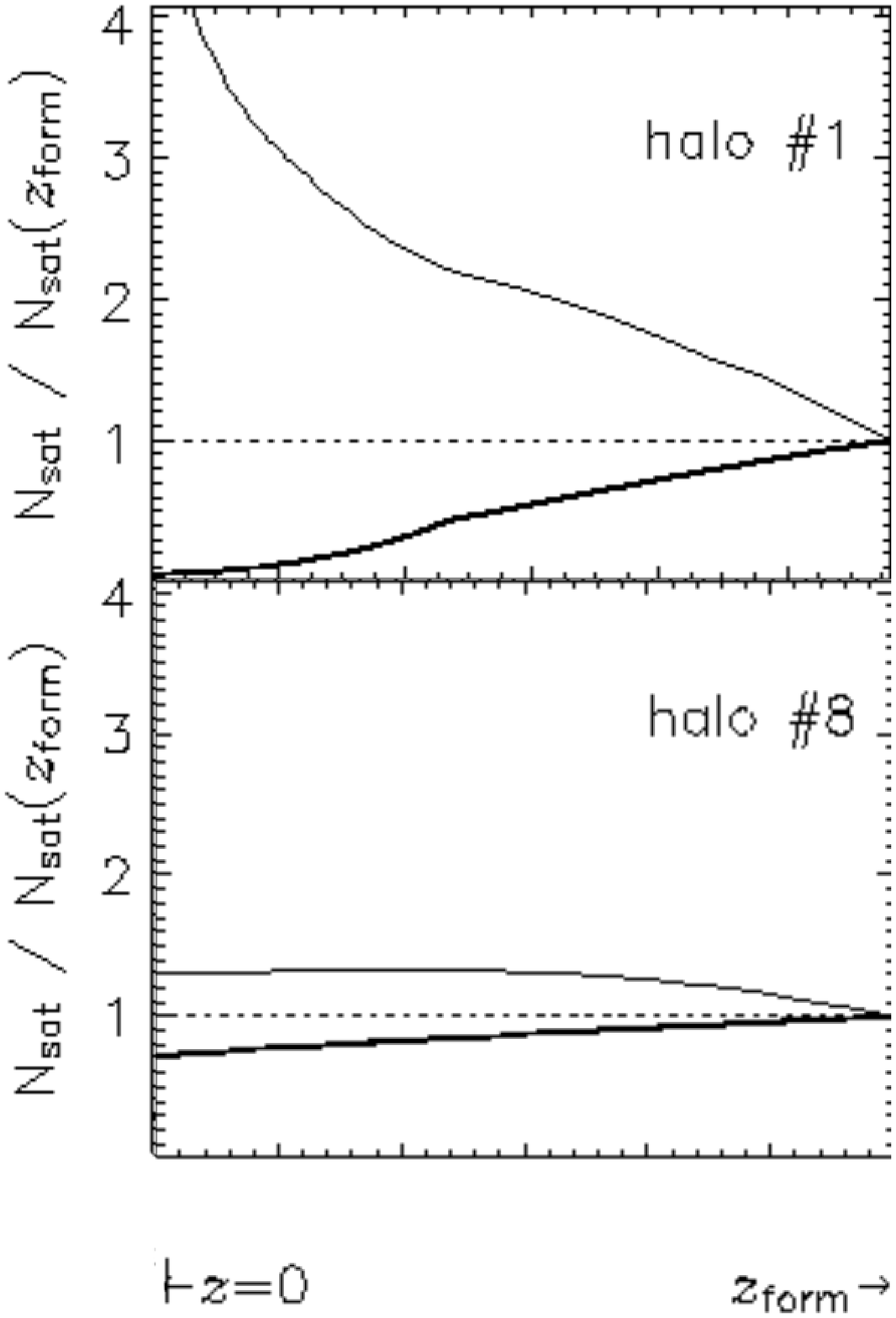}
\includegraphics{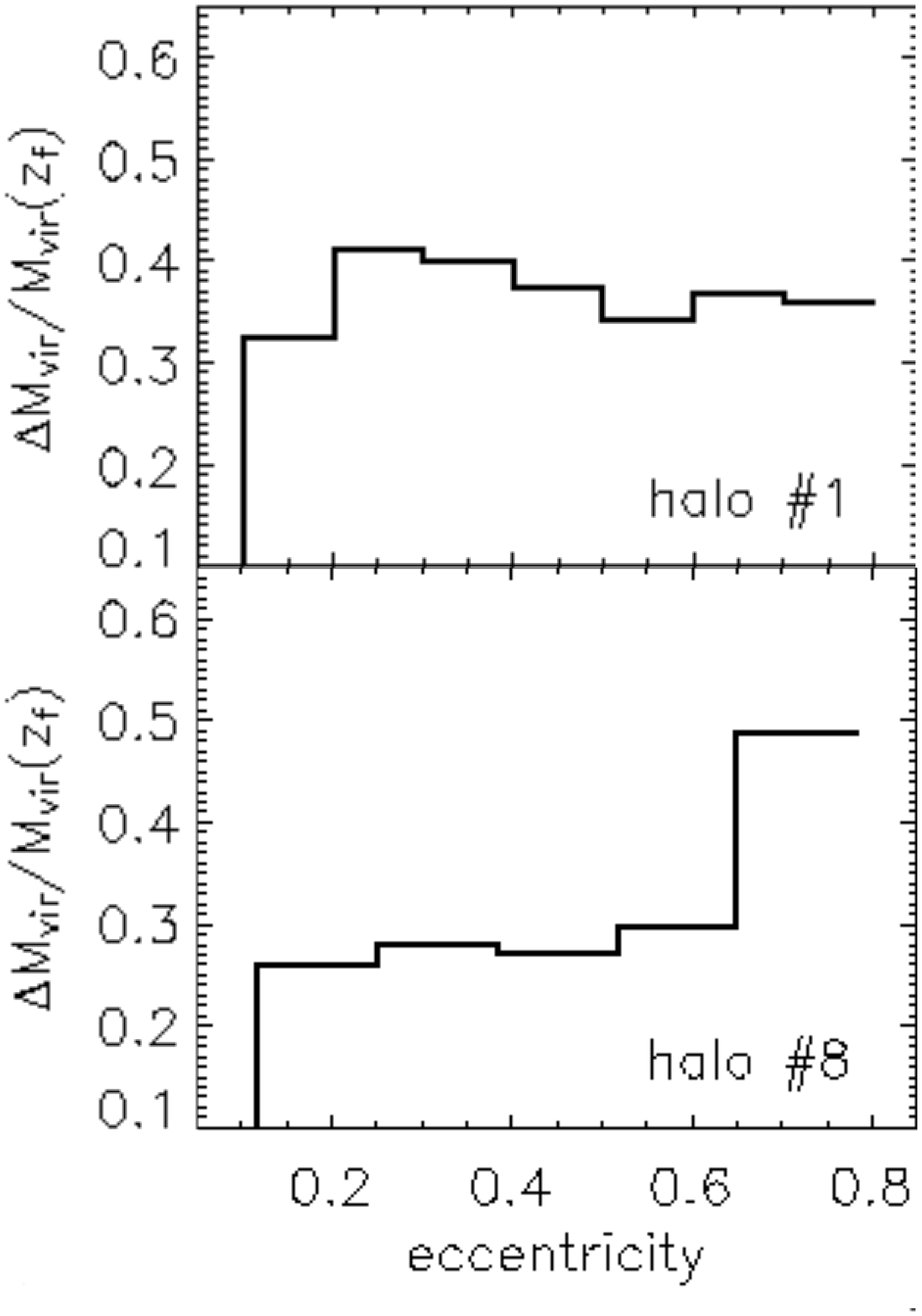}
   \caption{\footnotesize {(Left Panel) Number of satellites within a $2$\hMpc\ real co-ordinate sphere, as a function of redshift. Please note that the $x$-axis' are not the same range but rather covering the time from redshift $z=0$ to \zform\ for each individual halo. The thin line is the total number of satellites within that sphere whereas the thick line describes the disruption rate of the satellites within the sphere. (Right Panel) Mass loss per orbit for satellites that survived until redshift $z=0$ as a function of their orbital eccentricity $\epsilon$.}  }
\end{figure}

As the satellites orbit within the halo they experience dynamical friction causing them to sink towards the centre of the halo's potential, undergo tidal stripping, and gradually lose mass. For each satellite, its particle distribution becomes increasingly diffuse. Our criterion for calling a satellite "disrupted" is linked to its tidal radius; assuming that the average density of the satellite has to be at least a factor of three above the density of the host halo at the distance of the satellite we iteratively calculate the tidal radius and the mass $m( < r_{\rm tidal})$ enclosed by solving:

\begin{equation}\label{TidalRad}
	r_{\rm tidal} - \left( \frac{m( < r_{\rm tidal})}{3M( < D)} \right)^{\frac{1}{3}} D = 0\ ,
\end{equation}

The distance D of the satellite to the host and $M(<D)$, the mass of the host halo interior to D, are fixed.  As soon as there are fewer than 15 particles within the tidal radius we classify the satellite as being disrupted. Note that we are unable to separate numerical resolution disruption and real physical disruption of a satellite. It is not clear if we had infinite mass resolution that the satellite would still actually survive.

In Figure 2, the right panel shows the difference of the satellite's mass at \zform and z=0 divided by its mass at \zform and the number of orbits as a function of eccentricity of the satellite orbit. Thus as the satellite does get disrupted in general it losses at least 30 percent of its mass each orbit. Further, for seven of the eight halos, satellites on high eccentricity orbits are losing more mass (per orbit) than spherical ones, (such as halo \# 8). However, this is not strictly true for all environments; e.g. in halo \# 1, satellites on spherical orbits lose more mass - in this case the satellite orbits are on average closer to the halo core. 


\section{Properties of the Satellites in the Halos}

Because of the high temporal resolution of the outputs of our simulations ($\Delta t \sim 0.2$ Gyr) we are able to track accurately the orbital and other characteristics of the satellites. We present here a few preliminary results:

\begin{itemize}

\item For each of the 8 halos the mean eccentricity was consistently around 0.6, where eccentricity was measured as $ e = 1 - \frac{{\rm pericentre}}{{\rm apocentre}}$. 

\item For each of the 8 halos the mean pericentre distance was approximately 30 percent of the host virial radius.

\item The apocentres of the satellite orbits are aligned with the principle axis of the host galaxy cluster.

\item In general a satellite loses at least 30 percent of its mass each orbit. Further, for the majority of the eight halos, satellites on high eccentricity orbits are losing more mass (per orbit) because they penetrate deeper into the host's potential.

\item Satellite-Satellite interactions are not negligible.

\end{itemize}

A fuller and quantitative analysis is being prepared (Gill, Knebe~\& Gibson 2003). There, we address additional characteristics concerning satellite evolution: e.g. how does a satellite actually react to its tidal disruption?


\section{Acknowledgments}
SPDG would like to thank the Astronomical Society of Australia (ASA) and the conference organizers for assistance in attending the conference and the ongoing support of the Australian Research Council.


\end{document}